\documentstyle[preprint,aps]{revtex}
\newcommand{\lsim}{\mathrel{\mathop{\kern 0pt \rlap
  {\raise.2ex\hbox{$<$}}}
  \lower.9ex\hbox{\kern-.190em $\sim$}}}
\newcommand{\gsim}{\mathrel{\mathop{\kern 0pt \rlap
  {\raise.2ex\hbox{$>$}}}
  \lower.9ex\hbox{\kern-.190em $\sim$}}}

\newcommand{\beq}     {\begin{equation}}
\newcommand{\eeq}     {\end{equation}}

\newcommand{\es}      {\epsilon}

\newcommand{\lm}      {\lambda}
\newcommand{\M}       {{\mathcal M}}
\newcommand{\no}      {\nonumber}

\begin{document}

\draft

\preprint{
\vbox{\hbox{ASITP-2001-002}
      \hbox{KIAS-P01007}
      \hbox{hep-ph/0102246}}
}

\title{
A Model for Neutrino Warm Dark Matter \\
and Neutrino Oscillations
}

\author{
Chun Liu$^{\:a}$ and Jeonghyeon Song$^{\:b}$
}

\vspace{1.5cm}

\address{
$^a$Institute of Theoretical Physics, Chinese Academy of Sciences,\\
P.O. Box 2735, Beijing 100080, China\\
$^b$Korea Institute for Advanced Study, 207-43 Cheongryangri-dong,
Dongdaemun-gu,\\
Seoul 130-012, Korea
}

\maketitle

\thispagestyle{empty}

\setcounter{page}{1}

\begin{abstract}
The muon- and tau-neutrinos with the mass in the keV range,
which are
allowed in a low reheating temperature
cosmology, can compose the warm
dark matter of the universe.
A model of four light neutrinos including
the keV scale
$\nu_\mu$ and $\nu_\tau$ is studied,
which combines the
seesaw mechanism and the Abelian flavor symmetry.
 The atmospheric
neutrino anomaly is due to the $\nu_\mu-\nu_\tau$
oscillation.  The
solar neutrino problem is answered by
the oscillation into the light
sterile neutrino,
where the SMA, LMA, and LOW-QVO solutions
can be
accommodated in our scenario.
\end{abstract}



\newpage

Cosmological studies show increasing evidences
that the dark matter
is
in fact warm, neither cold nor hot,
so as to explain the observed
structure and behavior of the expanding universe \cite{wdm}.
 In
particle physics, as the candidates of the warm dark matter
(WDM),
gravitinos or sterile neutrinos with masses in the keV range have
been
proposed.  In this paper, we consider the possibility that the
ordinary active neutrinos are the WDM.

Experiments have provided various constraints on the neutrino mass
spectrum.  From the direct experimental search,
we have
$m_{\nu_1}\lsim 2.5$ eV, $m_{\nu_2}\lsim 170$ keV and
$m_{\nu_3}\lsim 15.5$ MeV \cite{pdg}.  More information comes from the
neutrino oscillation experiments.  The Super-Kamiokande (Super-K) data
for the atmospheric neutrino anomaly suggest that the $\nu_\mu$ is
maximally mixed with $\nu_x$ ($x\neq e$) with
$\Delta m_{\mu x}^2\simeq 2.2 \times 10^{-3}$ eV$^2$ \cite{superk}.
And the $x=\tau$ case is strongly favored\cite{superk1}.  The solar
neutrino deficit \cite{solar} may imply the oscillation of $\nu_e$
into $\nu_y$.  There are several allowed parameter regions.  For
example, if the $\nu_y$ is a sterile neutrino,
the currently favored
solution is the
small mixing angle (SMA) with
$\Delta m_{ey}^2 \simeq 5 \times 10^{-6}$ eV$^2$ and
$\tan^2\theta_{ey} \simeq 10^{-3}$.
 If one experiment among Super-K,
Ga and Cl is removed from the data analysis,
the large mixing angle
(LMA) solution with $\Delta m^2 \simeq 10^{-5}-10^{-4}$ eV$^2$
and the low-mass and quasi-vacuum oscillation (LOW-QVO)
solution with
$\Delta m^2 \simeq 10^{-10}-10^{-7}$ eV$^2$
are also allowed.  The
LSND
experiment has reported positive appearance results of
$\bar{\nu}_\mu \to \bar{\nu}_e$ oscillations, which implies
$\Delta m_{e\mu}^2\simeq 1$ eV$^2$ and
$\sin^2(2\theta_{e\mu})\simeq 10^{-2}$ \cite{lsnd}.
However, a
large part of its parameter space is excluded by the null
results of
the KARMEN data \cite{karmen}.

One of the most stringent constraints on the neutrino mass comes from
the cosmological consideration to avoid over-closing the universe.  In
the standard cosmology \cite{cos}, the stable neutrinos should be no
heavier than $20$ eV.  Recently it has been shown that if the
reheating temperature $T_{RH}$ is low, the densities of neutrinos can
become much less than usually assumed \cite{gkr,kks}.  The
cosmological constraint is very much relaxed.  Indeed, for
$1$ MeV $\lsim T_{RH} \lsim 3$ MeV, the abundance of tau- and
muon-neutrinos is \cite{gkrst}
\beq
\Omega_{\nu_\tau}h^2=\Omega_{\nu_\mu}h^2
=\left(\frac{m_\nu}{4~{\rm keV}}\right)
\left(\frac{ T_{RH}}{1~{\rm MeV} }\right)^3\,.
\eeq
This leads to Ref. \cite{gkrst} taking $\nu_\mu$ and $\nu_\tau$ as the
WDM.

The above WDM consideration makes the neutrino mass pattern quite
unique.  First the Super-K results constrain
the mass of both $\nu_\mu$
and $\nu_\tau$ around keV scale
\cite{superk1}.
Their mass-squared
difference
of order $10^{-3}$ eV$^2$
implies that the masses of
$\nu_\mu$ and $\nu_\tau$
are highly degenerate.
 In this case, the
solar neutrino problem
can only be understood by introducing
a light
sterile neutrino, $\nu_s$.
 The $\nu_e$ (and in the large mixing case
the $\nu_s$)
should be
lighter than eV order,
for the consistence with
the laboratory
experiments of $m_{\nu_e}$.
 The LSND result, however,
cannot be
compatible with the presence of keV scale muon-neutrinos.
Instead the mixing between the $\nu_\mu$ and $\nu_e$
should be very
small,
as expected from the keV scale mass of $\nu_\mu$
and the sub-eV
scale mass of $\nu_e$.

In this paper, a four light neutrino model is built to give the
above-mentioned neutrino mass pattern.  How to construct the mass
spectrum of four light neutrinos with hierarchies is a theoretically
challenging problem.  We extend a method proposed in Ref. \cite{ls}.
It simply incorporates the seesaw mechanism \cite{seesaw} with flavor
symmetry \cite{fs}.  Introducing three right-handed neutrinos and
assuming singularity in both the Dirac and Majorana mass matrices, the
neutrino mass matrix has approximately the following form:
\beq
\label{2}
{\cal M} = \left(
\begin{array}{cccccc}
0 & 0      & 0      & 0 & 0      & 0      \\
0 & 0      & 0      & 0 & m_{22} & m_{23} \\
0 & 0      & 0      & 0 & m_{32} & m_{33} \\
0 & 0      & 0      & 0 & 0      & 0      \\
0 & m_{22} & m_{32} & 0 & M_{22} & M_{23} \\
0 & m_{23} & m_{33} & 0 & M_{23} & M_{33} \\
\end{array}\right)\,.
\eeq
In this mass spectrum, there are two heavy neutrinos of masses
$\sim M$, two light neutrinos of masses $\sim m^2/M$, and two massless
neutrinos; three mass scales, $M$, $m^2/M$, and $0$, characterizes
this model.  Thus the four light neutrinos are naturally divided into
two pairs with a mass gap of $\sim m^2/M$, and each pair consists of
two degenerate mass eigenstates.

Abelian flavor symmetry \cite{fs} can be used to generate such a
neutrino mass matrix.  Furthermore this symmetry should also provide
the maximal mixing for the atmosphere neutrino anomaly which is not
guaranteed in the form of ${\mathcal M}$ in Eq.~(\ref{2}).  A softly
breaking of the symmetry is necessary to generate small masses for two
massless neutrinos and to lift the degeneracy in each pair.  In the
following discussion, supersymmetry is implied.  The flavor symmetry
is spontaneously broken by a vacuum expectation value (VEV) of an
electroweak singlet field $X$.  As long as the flavor charges balance
under the symmetry, the following interactions are allowed:
\beq
\label{3}
L_{\alpha} H N_{\beta}\left(
\frac{\langle X \rangle_{\rm VEV}}{\Lambda}\right)^{m_{\alpha\beta}}\,,
\quad
M N_\alpha N_\beta\left(
\frac{\langle X\rangle_{\rm VEV} }{\Lambda}\right)^{n_{\alpha\beta}}\,,
\eeq
where $L_\alpha$ ($\alpha=e$, $\mu$, $\tau$), $H$ and $N_\alpha$ denote
the lepton doublets, a Higgs field, and the right-handed neutrino
fields, respectively.  The $\Lambda$ is the flavor symmetry breaking
scale,
and the $m_{\alpha\beta}$ and $n_{\alpha\beta}$ are
non-negative integers,
required for the holomorphy of the superpotential.
The order parameter
for this new symmetry is defined by
\beq
\label{4}
\lm\equiv\frac{\langle X \rangle_{\rm VEV}}{\Lambda}\ll 1\,,
\eeq
which is
a typical order of
Cabbibo angle $\sim 0.1$

The assignment of the Abelian flavor charges relevant to Eq.~(\ref{2})
and to maximal $\nu_\mu-\nu_\tau$ mixing is assumed to be
\begin{eqnarray}
\label{assignment}
&& L_e(z+1/2),    ~~L_\mu(a),        ~~L_\tau(-a),     \\ \no
&& E^c_e(-z+9/2), ~~E^c_{\mu}(-a+3), ~~E^c_{\tau}(a+2),\\ \no
&& N_e(x-1/2),    ~~N_\mu(-a),       ~~N_\tau(a),   ~~X(-1),
\end{eqnarray}
with the positive integers $a$ and
$x(>a>1)$.
 The sign of the integer
$z$ is to be set later.
 The $E^c_{\alpha}$'s in Eq.~(\ref{assignment})
are the anti-particle
fields of the SU(2) singlet charged leptons.  In
order to obtain the
physical mixing angles of neutrinos, the charged
lepton mass matrix
should be simultaneously taken into account.  The
gauge and Higgs
bosons possess vanishing flavor charges.
 Note that the
flavor charges for the first generation
are half-integers while those
for the second and
third generations are integers expressed by a single
parameter $a$.  Compared to analogous analysis for three light
neutrino
scenario, the choice of the flavor charges here is more
limited.  One
tricky point is that one of the right-handed neutrino
masses is made to
be vanishingly small ($\ll m^2/M$).

The flavor charge assignment in Eq.~(\ref{assignment})
produces the
Dirac and Majorana mass matrices
of neutrinos as
\beq
\label{m-nu}
\M_D = m\left(
\begin{array}{ccc}
Y_{11} \lm^{x+z}   & 0 & 0                \\
0                  & 1 & Y_{23} \lm^{2 a} \\
0                  & 0 & 1
\end{array}
\right)\,,
\quad
\M_M = M\left(
\begin{array}{ccc}
\zeta_1 \lm^{2x-1} & 0 & 0                \\
0                  & 0 & 1                \\
0                  & 1 & \zeta_4 \lm^{2 a}
\end{array}
\right)\,,
\eeq
and the mass matrix of charged leptons as
\beq
\label{m-l}
\M_l = m \left(
\begin{array}{ccc}
\eta_{11}\lm^5     & 0              & 0                    \\
0                  & \eta_{22}\lm^3 & \eta_{23} \lm^{2a+2} \\
0                  & 0              & \eta_{33} \lm^2
\end{array}
\right)\,,
\eeq
where $Y$'s, $\zeta$'s and $\eta$'s are order one coefficients.  The
$\M_l$ and $\M_D$ are almost diagonal,  while the $\M_M$ is mainly
off-diagonal.  Therefore to leading order,
the mass matrix of four
light neutrinos in the
$(\nu_s,\nu_e,\nu_\mu,\nu_\tau)$
basis is
obtained as,
\beq
\label{M0}
\M_{\nu}^{(0)}\simeq\frac{m^2}{M}\left(
\begin{array}{cccc}
0 & 0  & 0  & 0  \\
0 & 0  & 0  & 0  \\
0 & 0  & 0  & -1 \\
0 & 0  & -1 & 0
\end{array}
\right)\,.
\eeq
The neutrino mass spectrum due to $\M_{\nu}^{(0)}$ is
\beq
m_{\nu_s}=m_{\nu_e}=0,\quad m_{\nu_\mu}=m_{\nu_\tau}
=\frac{m^2}{ M},\quad
\sin\theta_{\mu\tau}=\frac{1}{\sqrt{2}}\,.
\end{equation}

It is to be noted that the maximal mixing
between the $\nu_\mu$ and
$\nu_\tau$ results from the
flavor symmetry.
 The requirement of keV
scale $\nu_\mu$ and $\nu_\tau$,
i.e.,
\beq
\label{keV}
\frac{m^2 }{ M} \sim 1 \, {\rm keV}\,,
\eeq
is achieved via the seesaw mechanism
with $m \simeq 300$ GeV
and
$M \simeq 10^{11}$ GeV.
 The $\es$, defined by the ratio $m/M$,
is
then of order $10^{-8.5}$.

First let us examine the mass spectrum
in the charged lepton sector.
The eigenvalues of the $\M_l$ are of order
$\lm^5 m$, $\lm^3 m$, and
$\lm^2 m$,
which yields appropriate mass
scales of the charged
leptons.
 The $\M_l $ is diagonalized by
\beq
R_l^L \M_l \; R_l^{R\dagger} =
{\rm Diag}\; (m_e, m_\mu, m_\tau) \,,
\eeq
where the $R_l^L$ diagonalizes the hermitian
mass-squared matrix
$\M_l \M_l^\dagger $.

In the neutrino sector the mass matrix of four light neutrinos can be
obtained by the method described in Ref.~\cite{ssm2}.
 We finally
have the following symmetric mass matrix
of four light neutrinos in
the
$(\nu_s,\nu_e,\nu_\mu,\nu_\tau)$
basis with
$\lm^x/\es \equiv \lm^\beta$:
\beq
\label{15}
\M_{\nu}=\frac{m^2}{M}\left(
\begin{array}{cccc}
\lm^{2\beta-1} & \lm^{\beta+z} & 0       & 0  \\
\lm^{\beta+z}  & 0             & 0       & 0  \\
0              & 0             &\lm^{2a} & -1 \\
0              & 0             & -1      & 0
\end{array}
\right)\,,
\eeq
where the charged lepton mixing effects are
incorporated so that the
charged lepton
fields have been rotated into
mass eigenstates.
In
Eq.~(\ref{15}) the combinations of order one parameters
such as
$Y_{ij}$, $\eta_i$, and $\zeta_i$
are omitted, for simplicity.

In our scenario, there is no mixing between
the two pairs due to our assignment
of half-integer charges for the first generation,
and integer charges for the second and third generations.
 The
absence of the $\nu_e-\nu_\mu$ mixing
is compatible with the KARMEN
data,
predicts null results in future laboratory searching for the
$\nu_e-\nu_\mu$ or $\nu_e-\nu_\tau$ mixing,
and has no influence on
astrophysical processes
\cite{future}.
Then the
${\mathcal M}_\nu$
is
automatically factorized out
into $M_{\nu_s-\nu_e}$ and
$M_{\nu_\mu-\nu_\tau}$.

The requirement of Super-K data determines the value of $a$ as
\beq
\Delta^2 m_{23} \simeq \lm^{2 a} \left(\frac{m^2}{ M}\right)^2
\simeq 3 \times 10^{-3}\,{\rm eV}^2,~~~{\rm for}~\, a=4 \,.
\eeq

The effective mass matrix of $\nu_e$ and $\nu_s$
in the unit of eV
can be written by
\beq\label{Mse}
M_{\nu_s-\nu_e} \simeq {\rm eV}\left(
\begin{array}{cc}
\lm^{2\beta-4}  & \lm^{\beta+z-3}     \\
\lm^{\beta+z-3} & 0
\end{array}
\right)\,.
\eeq
The case of $\beta-1 < z$ $(\beta-1>z)$
corresponds to the small
(large) mixing angle solution
for the solar neutrino problem.
Some possible solutions
are listed in the following table.
Note that in the SMA case, the $m_{\nu_e}$
is the smaller mass since the $M_{\nu_s-\nu_e}$ in
Eq.~(\ref{Mse}) is in the $(\nu_s,\nu_e)$ basis.

\vspace{1cm}
\begin{center}
\begin{tabular}{|c|r|c|c|c|c|}
\hline
\quad Solution \quad & \quad $z$ \quad
& \quad $\beta~(x)$ \quad
& \quad $m_{\nu_e}$ (eV)
\quad & \quad $\Delta m^2_{\odot}$ (eV$^2$)
\quad
& \quad $\sin^2 2\theta_\odot$
\\ \hline
\quad SMA \quad & \quad 4 \quad & \quad 3.5~(12) \quad
& \quad $10^{-6}$
\quad & \quad $10^{-6}$ \quad
& \quad $10^{-3}$ \\
\quad LMA \quad & \quad 1 \quad & \quad 3.5~(12) \quad
& \quad $3\times 10^{-2}$
\quad & \quad $3 \times 10^{-5}$ \quad
& \quad $1$ \\
\quad LOW-QVO \quad & \quad 1 \quad & \quad 4.5~(13) \quad
& \quad $3\times 10^{-3}$
\quad & \quad $3\times 10^{-8}$ \quad
& \quad $1$ \\
& \quad 2 \quad & \quad 4.5~(13) \quad
& \quad $3\times 10^{-4}$
\quad & \quad $3\times 10^{-9}$ \quad
& \quad $1$ \\
& \quad 3 \quad & \quad 4.5~(13) \quad
& \quad $3\times 10^{-5}$
\quad & \quad $3\times 10^{-10}$ \quad
& \quad $1$ \\
& \quad 0 \quad & \quad 5.5~(14) \quad
& \quad $3\times 10^{-3}$
\quad & \quad $3\times 10^{-10}$ \quad
& \quad $1$ \\
& \quad $-1$ \quad & \quad 5.5~(14) \quad
& \quad $3\times 10^{-2}$
\quad & \quad $3\times 10^{-9}$ \quad
& \quad $1$ \\
& \quad $-2$ \quad & \quad 5.5~(14) \quad & \quad 0.3
\quad
& \quad $3\times 10^{-8}$ \quad & \quad $1$ \\
\hline
\end{tabular}
\end{center}
\vspace{1cm}

It is to be checked in the SNO \cite{sno}
and KamLAND experiments.
For one of the LOW-QVO solution, the mass of the electron neutrino
is just an order
of magnitude lower than the current experiment
limit.
 It is possible
to probe this value in future experiments
directly.
 In addition, it may have observable effect
in the cosmic
microwave background anisotropies.
 The $\nu_e$ and $\nu_s$ compose
a Dirac neutrino which does not
result in any observable neutrinoless
double $\beta$ decay of relevant
next generation experiments.

In summary, it would be simple if the muon- and tau-neutrinos
are
just the WDM.
 We have presented a neutrino model which generates
keV scale
$\nu_\mu$ and $\nu_\tau$ and meanwhile provides the
neutrino
oscillation solutions for the solar and atmospheric
neutrino experiments.
 It combines the seesaw mechanism and the
Frogatt-Nielsen
mechanism. The key point is that the neutrino mass
matrix has a
singular form.
 A light sterile neutrino is obtained
due to the flavor
symmetry.

Finally some remarks should be mentioned.
\begin{itemize}
\item
Compared to the model
in Ref. \cite{ls},
the light neutrino spectrum is
similar to the
ordinary $2+2$ light neutrino scheme \cite{four}.  But
the two
neutrino pairs are more widely separated (by keV).
 For the solar
neutrino problem,
the SMA, LMA or LOW-QVO oscillation
solutions into light sterile neutrinos
can be
accommodated in our scenario.
 This is achieved by taking the
U(1) flavor charges
to be half-integers for the first generation and
to be integers for the second and third generations.  Of course, the
LSND results cannot be relevant in this model.
 But our results are
compatible with the KARMEN data.
\item
The LMA and LOW-QVO solutions
have been obtained in fact due to
the singular seesaw mechanism.  The singular seesaw mechanism was
applied to the atmospheric neutrino anomaly in Ref. \cite{ssm2}.
However, the current Super-K data do not favor the $\nu_\mu-\nu_s$
oscillation.  What we have done in Ref. \cite{ls} and in this paper
for LMA and LOW-QVO scenarios are to obtain a singular seesaw
mechanism for the solar neutrinos.
\item
It seems a drawback to introduce
both the seesaw and the
Frogatt-Nielsen
mechanism to make neutrinos light.
However the
seesaw mechanism itself
cannot predict the neutrino flavor structure,
especially the maximal
$\nu_\mu-\nu_\tau$ mixing.
 Another underlying
motivation is to make the
four light neutrino scenario easier to be
understood in the framework of
grand unification theories, which will
be studied further.
\item Other interesting aspects of keV neutrinos were discussed before.
Such neutrinos may be responsible for the large velocity of pulsars
\cite{19}.  Because the neutrino densities in the early universe are
much smaller than that assumed in the standard cosmology, the
cosmological constraints for neutrinos from the big bang
nucleosynthesis is different from that in the standard one \cite{20}.
This needs more detailed study.  In addition to the keV active
neutrinos, certain amount of some other keV sterile neutrinos \cite{21}
may also contribute to the WDM.
\end{itemize}

\acknowledgments
C.L. was supported in part by the National Natural Science Foundation of
China with grant no. 10047005.

\end{document}